\documentstyle[pre,aps,epsf,multicol]{revtex}
\begin{document}

\title{Critical behavior of the two dimensional $2A\to 3A$, $4A\to\emptyset$
binary system}
\author{G\'eza \'Odor}
\address{Research Institute for Technical Physics and Materials Science, \\
H-1525 Budapest, P.O.Box 49, Hungary}    
\maketitle

\begin{abstract}
The phase transitions of the recently introduced $2A\to 3A$, $4A\to\emptyset$ 
reaction-diffusion model (G.\'Odor, PRE 69 036112 (2004)) are explored in 
two dimensions.
This model exhibits site occupation restriction and explicit diffusion
of isolated particles. A reentrant phase diagram in the diffusion - creation
rate space is confirmed in agreement with cluster mean-field and 
one-dimensional results. For strong diffusion a mean-field transition 
can be observed at zero branching rate characterized by $\alpha=1/3$ density
decay exponent. In contrast with this for weak diffusion the effective
$2A\to 3A\to 4A\to\emptyset$ reaction becomes relevant and the mean-field 
transition of the $2A\to 3A$, $2A\to\emptyset$ model characterized by 
$\alpha=1/2$ also appears for non-zero branching rates. 
\end{abstract}
\begin{multicols}{2}

\section{Introduction}

The classification of universality classes of nonequilibrium systems is one
of the most important tasks of statistical physics \cite{Uweof,dok}.
Many of the known systems can be mapped onto some reaction-diffusion type of
models, the behavior of them are the studied intensively in the past decades
\cite{DickMar,Hin2000}. In these systems particle (A) creation, 
annihilation and diffusion processes compete and by tuning the control 
parameters phase transition may occur from an active steady state to an 
inactive, absorbing state of zero density. For a long time only the critical 
``directed percolation'' (DP) type of universality class has been known 
\cite{DPuni}.
Later other classes were discovered related to certain conservation laws or
symmetries \cite{Gras84,NEKIM,Cardy-Tauber,rossi}, to long-range interactions 
\cite{MF94,JOWH99,HH99}, to boundary conditions
\cite{FHL01,Turban,HaOd98,PCP} or to disorder 
\cite{MoreiraDickman96,Noest,WACH98,CGM98,HIV02}.
These findings are all in agreement with the concepts of universality in 
equilibrium systems. 

An extraordinary family of models has triggered a long debate among specialists
recently \cite{GrasBP,HT97,Carlon99,Hayepcpd,Odo00,HayeDP-ARW,coagcikk,MHUS,HH,binary,OSC02,multipcpd,PK66,pcpd2cikk,KC0208497,DM0207720,PHK02,brazcikk,2340cikk,NP01,HH04of}. 
The common behavior of these models is that for a particle production and
annihilation at least two particles are needed (henceforth they are called 
binary systems) and these reactions compete with the diffusion of isolated 
particles. Since for reactions at least a pair is needed while isolated 
particles can diffuse only these models can also be regarded as coupled 
systems \cite{HayeDP-ARW}. The representative of this class is the so called
diffusive pair contact process (PCPD) with reactions 
$2A\to 3A$, $2A\to\emptyset$ \cite{Carlon99}. 
The binary nature was found to be relevant in case of reactions of
multi-species \cite{multipcpd} too.
  
The critical behavior of such models has been found to be different from all 
previously known classes (however there is still an ongoing debate on the 
precise values of critical exponents). The lack of symmetries, conservation 
laws etc. have been motivating skepticism against the existence of a non-DP 
class transition and recently some studies suggested DP class behavior 
with extremely strong correction to scalings \cite{HinDP1,HinDP2,BarkCar}. 
Field theoretical analysis  \cite{HT97} on the other hand indicate that the 
absence of the mass term corresponding to direct channel to the absorbing 
state ($A\to\emptyset$) should be responsible for this ``anomalous'' 
behavior with 
respect to expectations based on equilibrium statistical physics. 
There is an other important difference between binary systems and DP: 
there is no rapidity symmetry
\begin{equation}
\phi(x,t) \to -\psi(x,-t)  \ \ , \ \  \psi(x,t) \to -\phi(x,-t) \ \ .
\label{DPsymeq}
\end{equation}
between the field ($\phi$) and the response field ($\psi$) variables
in the corresponding field theoretical description contrary the case
of the DP process.
Furthermore the lack of this relation is not the consequence of a symmetry
breaking field of some boundary (like the $t=0$ boundary with long-ranged 
correlated order parameter field in case of pair contact process (PCP) 
\cite{PCP}) or some 
disorder, but it is not there in the definition of these homogeneous, 
binary systems \footnote{Noh et al. claim that in their generalized PCPD 
model a long-range memory is generated by the diffusing isolated particles 
\cite{NP01}}.

Another odd feature is that bosonic (site unrestricted) and site restricted
versions of these models show completely different behavior. While site 
restricted models investigated numerically exhibit the above continuous 
phase transition, the bosonic versions do not have steady state, but above 
an abrupt transition the density of particles diverges quickly 
\cite{HT97,OdMe02}. The field theoretical renormalization 
group (RG) analysis \cite{HT97} predicts an upper critical dimension 
$d_c=2$, with logarithmic corrections at $d=2$ for this class (PCPD).
Simulations \cite{OSC02} have confirmed the mean-field scaling in two 
dimension in case of the $2A\to 4A$, $2A\to\emptyset$ binary production 
model.

The site mean-field solution of general,
\begin{equation}
n A \stackrel{\sigma}{\to} (n+k)A,
\qquad m A \stackrel{\lambda}{\to} (m-l) A, \label{genreactions}
\end{equation}
models (with $n>1$, $m>1$, $k>0$, $l>0$ and $m-l\ge 0$) 
resulted in a series of different universality classes 
{\bf depending on $n$ and $m$} \cite{tripcikk}. This shows that 
above $d_c$ {\bf $n$ and $m$ are relevant parameters} determining the
type of continuous phase transitions. 
In particular for the $n=m$ symmetrical case the density of particles
above the critical point ($\sigma_c > 0$) scales as
\begin{equation}
\rho \propto |\sigma-\sigma_c|^{\beta}, \label{betascale}
\end{equation}
with $\beta^{MF}=1$, while at the critical point it decays as
\begin{equation}
\rho \propto t^{-\alpha} \ , \label{alphascale}
\end{equation}
with $\alpha^{MF}=\beta^{MF}/\nu^{MF}_{||}=1/n$ \cite{PHK02,tripcikk}
(here "MF" denotes mean-field value).
On the other hand for the $n<m$ asymmetric case continuous phase 
transitions at zero branching rate $\sigma_c=0$ occur with
\begin{equation}
\beta^{MF}=1/(m-n), \quad \alpha^{MF}=1/(m-1) \label{asymmfscal}
\end{equation}
For $n>m$ the mean-field solution provides first order
transition.

By going beyond site mean-field approximations it turns out that the
above classification is not completely satisfying. 
In a previous paper \cite{2340cikk} I investigated the $2A\to 3A$, 
$4A\to\emptyset$ model by cluster mean-field approximations and simulations 
in 1d and showed that the {\bf diffusion} plays an important role: 
it introduces a different critical point besides the one at $\sigma=0$ 
branching rate with eq. (\ref{asymmfscal}) exponents.
The non-trivial critical point, obtained for low diffusion rate exhibits the 
universal behavior of the $2A\to 3A$, $2A\to\emptyset$ (PCPD) model owing to 
the generation of the effective $2A\to\emptyset$ reaction via: 
$2A\to 3A\to 4A\to\emptyset$ \cite{Ccom}.
\begin{figure}
\begin{center}
\epsfxsize=70mm
\centerline{\epsffile{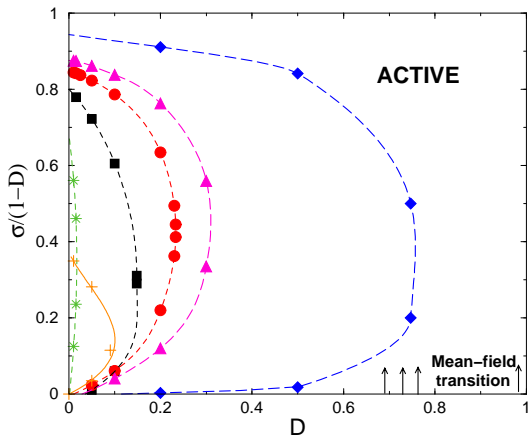}}
\caption{Phase diagram of the $2A\to 3A\to 4A\to\emptyset$ model. 
Stars correspond to $N=2$, boxes to $N=3$, bullets to $N=4$ and 
triangles to $N=5$ cluster mean-field approximations. 
Diamonds denote 1d, $+$ signs 2d simulation data, where PCPD class 
transitions are found. The lines serve to guide the eye. 
At the $\sigma=0$ line asymmetric, eq. (\ref{asymmfscal}) type 
mean-field transition occurs.}
\label{pd}
\end{center}
\end{figure}
In this work I continue the study of this model in 2d and show, that
similar phase transition structure and critical behavior can be obtained.
This is somewhat surprising, since one may expect that the diffusion is less
relevant in higher dimensions due to its short interaction range. 
A very recent study using exact methods \cite{PS02} showed that the particle
density fluctuation and density correlation function are diffusion dependent
in the bosonic PCPD model for $d>2$. In this work I give numerical evidence 
for diffusion dependence in a site restricted, binary model in $d=2$.

\section{The $2A\to 3A$, $4A\to\emptyset$ model}

This binary production reaction-diffusion model is defined by the rules:
\begin{eqnarray}
2A \rightarrow 3A  \qquad {\rm with \ rate} \, & \ \ \sigma \nonumber \\
4A \rightarrow \emptyset \qquad  {\rm with \ rate}\,  
& \lambda = 1-\sigma \nonumber \\
A\emptyset \leftrightarrow \emptyset A \qquad {\rm with \ rate}\,  & D \ .
\label{DynamicRules}
\end{eqnarray}
Here $D$ denotes the diffusion probability and $\sigma$ is the production
probability of the particles. 
The site occupancy is restricted to 0 or 1 particle.
In \cite{2340cikk} the cluster mean-field approximations were determined on
1d lattices for $N=1,2,..5$ cluster sizes. The corresponding reentrant phase
diagram is shown on Fig. \ref{pd}. Although cluster mean-field approximations 
based on $d>1$ lattices may result in transition points at other locations, 
the universal features are expected to be the same.
Therefore I compare the simulation results with this approximation.
\begin{figure}
\begin{center}
\epsfxsize=70mm
\centerline{\epsffile{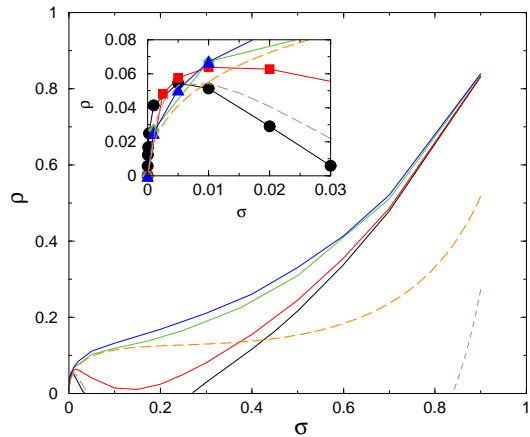}}
\caption{Simulation results for the steady state density at diffusions
$D=0.5$, $0.35$, $0.1$, $0.05$ (solid lines from top to bottom) 
and $N=5$ level cluster mean-field approximation data for $D=0.5$,
$0.05$ (dashed lines from top to bottom). The insert shows the region 
near $\sigma=0$ magnified.}
\label{rs}
\end{center}
\end{figure}
.

\subsection{Simulation results}

I performed simulations in two dimensions in $L=1-7\times 10^3$ linear
sized systems with periodic boundary conditions. The simulations were 
started from fully occupied lattices.
One elementary Monte Carlo step consists of the following processes.
A particle and a number $x_1 \in (0,1)$ are selected randomly; 
if $x_1 < D$ a site exchange is attempted with one of 
the randomly selected empty nearest neighbors (nn). The time is updated
by $1/n$, where $n$ is the total number of particles.
A particle and a number $x_2 \in (0,1)$ are selected randomly.
If $x_2 < \sigma$ and if the number of nn particles is 1 or 2 or 3,
one new particle is created at an empty site selected randomly.
If $x_2 \ge \sigma$ and the number of nn particles is greater than 2
four randomly selected neighboring particles are removed.
The time ($t$) is updated by $1/n$ again.
The density of particles was followed up to $t_{max} \le 10^7$ 
Monte Carlo steps (throughout the whole paper the time is 
measured by Monte Carlo steps (MCS)).

As one can see on Fig.\ref{rs} simulation data and the 5-point 
approximations fit qualitatively well. In both cases for weak diffusion
rates (for $D < \sim 0.1$ in 2d simulations) reentrant phase transitions 
occur with $\sigma_c > 0$,
while for strong diffusions a single phase transition at $\sigma_c=0$ 
branching rate can be found. The transition lines of the cluster mean-field 
approximations do not converge towards the simulation line as in 1d 
(see Fig.\ref{pd}), but the 2d MC curve occurs at lower diffusions.
But this is not surprising, since the cluster mean-field calculations are 
performed on 1d lattices.
\begin{figure}
\epsfxsize=70mm
\epsffile{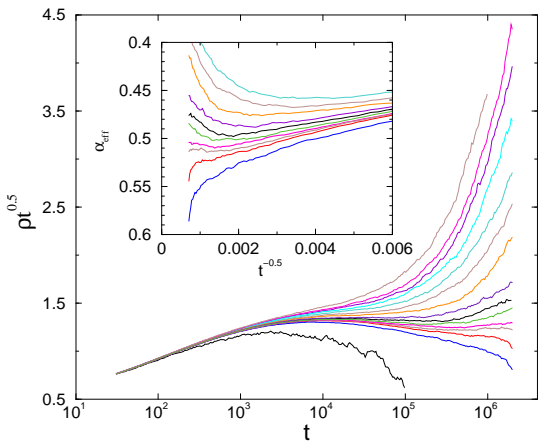}
\caption{Density decay times $t^{0.5}$ in the two-dimensional $2A\to 3A$,
$4A\to\emptyset$ model at $D=0.05$.
Different curves correspond to $\sigma=0.2715$, $0.2708$, $0.2704$, $0.27$,
$0.2695$, $0.269$, $0.2685$, $0.268$, $0.2677$, $0.2675$, $0.2673$, $0.26715$,
$0.267$, $0.2665$, $0.26$ (top to bottom). The insert shows the corresponding
local slopes.}
\label{sl5}
\end{figure}
I explored the scaling behavior in more detail at $D=0.05$ diffusion near
the rightmost transition of Fig. \ref{rs} (at $\sigma \sim 0.27$).
By approaching $\sigma_c$ from the active phase the $\rho(t) t^{1/2}$ 
curves bend down rapidly for long times (beyond $\sim 10^6$ MCS).
However this proved to be a finite size effect : the break-down 
of the density curves can be eliminated by increasing $L$.
The largest system I could simulate had a linear size $L=7000$. 
In this case no rapid and premature curvatures was observed for 
$t < 2\times 10^6$ MCS.
As one can see on Fig.\ref{sl5} for $\sigma>0.2673$ all curves veer up,
while for $\sigma < 0.2673$ they veer down. A clear straight line --
indicating scaling with the expected logarithmic correction -- can not be seen
clearly. Even the $\sigma=0.2673$ curve shows some up and down curvatures
in the last decade of the simulations.
However as can be seen on the local slopes figure (see insert of Fig.\ref{sl5})
defined as:
\begin{equation}
\alpha_{eff}(t) = {- \ln \left[ \rho(t) / \rho(t/m) \right] 
\over \ln(m)} \ , \label{slopes}
\end{equation}
(where I used $m=2$) the transition is around the expected 
mean-field value of the PCPD class: $\alpha=0.5$ \cite{HT97,tripcikk}.
Other curves exhibit strong curvatures for long times,
i.e. for $\sigma > 0.2673$ they veer up (active phase), while for
$\sigma < 0.2673$ they veer down (absorbing phase).

\begin{figure}
\begin{center}
\epsfxsize=70mm
\centerline{\epsffile{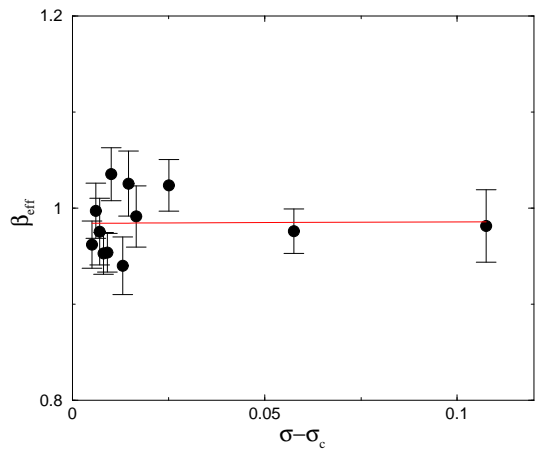}}
\caption{$\beta_{eff}$ as the function of $\sigma-\sigma_c$ in the
two dimensional $2A\to 3A$, $4A\to\emptyset$ model near the 
$\sigma_c=0.2673$ critical point for $D=0.05$.
The solid line shows a linear fitting.}
\label{betaeff_05}
\end{center}
\end{figure}

The steady state density in the active phase near the critical
phase transition point is expected to scale as 
$\rho(\infty)\propto|\sigma-\sigma_c|^{\beta}$.
Using the local slopes method one can get a precise estimate for
$\beta$ and see the corrections to scaling
\begin{equation}
\beta_{eff}(p_i) = \frac {\ln \rho(\infty,\sigma_i) -
\ln \rho(\infty,\sigma_{i-1})} {\ln(\sigma_i) - \ln(\sigma_{i-1})} \ .
\label{beff}
\end{equation}
The steady state behavior at the $\sigma_c > 0$ transition for $D=0.05$ 
was investigated using $\sigma_c=0.2673(2)$ from the density decay analysis. 
Here the local slopes tend to 
$\beta_{eff}=0.98(2)$ without showing any relevant correction to scaling
(see Fig.\ref{betaeff_05}). This agrees with the mean-field value of
the PCPD model again \cite{HT97,tripcikk}.
One may expect the same kind of transition all along the $\sigma_c > 0$ 
transition line. Indeed simulations showed that the density decays in a 
similar way at transitions with $D=0.01$, $0.05$, $0.09$.

To see the transition near $\sigma_c=0$ (horizontal axis on Fig.\ref{pd})
I determined the steady state value of $\rho(\infty,\sigma)$ for several 
$\sigma$-s at $D=0.05$ diffusion. 
The steady state density was determined by running the simulations
in the active phase near $\sigma = 0$, by averaging over $\sim 100$ 
samples in a time window following the level-off is achieved.
The smallest value I tested was $\sigma=10^{-5}$, when I had to go up to
$t=10^7$ MCS to reach a steady state (on a $L=2000$ sized system). 
By looking at the data it is quite obvious that the transition is at
$\sigma_c = 0$ as the cluster mean-field approximations predicted.

\begin{figure}
\begin{center}
\epsfxsize=70mm
\centerline{\epsffile{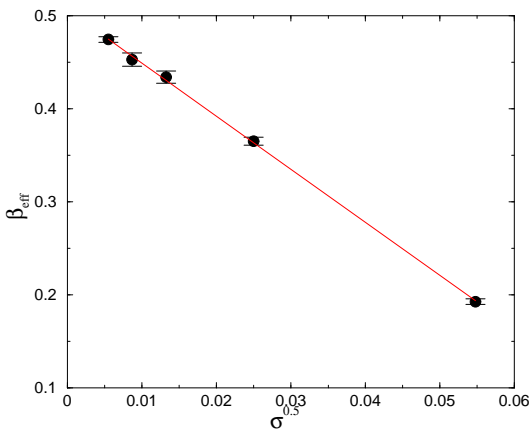}}
\caption{$\beta_{eff}$ as the function of $\sigma^{0.5}$ in the
two dimensional $2A\to 3A$, $4A\to\emptyset$ model near the $\sigma_c=0$
phase transition at $D=0.05$. The solid line shows a linear fitting.}
\label{beta_0}
\end{center}
\end{figure}

The effective order-parameter exponent (Fig.\ref{beta_0}) tends
to $\beta=0.505(5)$ as $\sigma\to 0$ corroborating the cluster
mean-field prediction: eq.(\ref{asymmfscal}). Assuming a correction to 
scaling of the form
\begin{equation}
\beta_{eff}=\beta - a t^{-\beta_1} \label{corsca}
\end{equation}
fitting results in $\beta_1=0.5$ as can be read-off from Fig.\ref{beta_0} . 

\section{Conclusions}

In conclusion I have investigated the ($D$-$\sigma$) phase diagram of 
the two dimensional $2A\to 3A$, $4A\to\emptyset$ model with site restriction 
and explicit particle diffusion. Extensive simulations gave numerical 
evidence that a reentrant phase diagram emerges as in one dimension
and predicted by cluster mean-field approximations \cite{2340cikk}.
This somewhat surprising results mean that diffusion plays relevant
role even in $d=2$ dimension. For high diffusion rates only a mean-field
transition at $\sigma=0$ branching rate can be found, while for low diffusion
an other transition type at $\sigma_c>0$ appears. This latter transition
shows the mean-field characteristics of the PCPD model because
the effective $2A\to\emptyset$ reaction (via $2A\to 3A\to 4A\to\emptyset$)
becomes relevant. The understanding of this diffusion dependence is a 
challenge for field theory.
Similar reentrant phase diagram has been observed in case of the unary
production, triplet annihilation model ($A\to 2A$, $3A\to\emptyset$)
\cite{Dicktrip} and in a variant of the NEKIM model \cite{nbarw2cikk}.
In all cases the diffusion competes with particle reaction processes,
and the bare parameters should somehow form renormalized reaction rates
which govern the evolution over long times and distances.
An interesting question is whether this scenario extends above $d=2$ 
dimensions as the cluster mean-field approximation predicts.
A very recent non-perturbative RG study \cite{CCD04} finds similar phase 
diagram in case of the $A\to 2A$, $2A\to\emptyset$ model for $d\ge 3$ 
dimensions. That work points out that non-perturbative effects arise
and there is a threshold $(\lambda/D)_{th}(d)$ above which DP, while below 
it a type (\ref{asymmfscal}) mean-field transition at $\sigma_c=0$ appears.

The simulations also showed that at the $\sigma_c>0$ transition the 
finite size effects and corrections to scaling are very strong.
I had to go up to $7000\times 7000$ sized systems and 
$t_{max}=2\times 10^6$ MCS to see the appearance of the expected 
mean-field scaling with exponents $\alpha=0.5$, $\beta=1$. 
Showing clear scaling for more than a decade with the predicted 
logarithmic corrections \cite{HT97} is beyond the scope of this study,
yet these simulation results for a 2d binary system are by far the 
largest scale ones published so far.
On contrary the scaling at the $\sigma_c=0$ critical point is clear 
with $\beta=0.505(5)$ and correction to scaling exponent $\beta_1=0.5$.
\bigskip 

{\bf Acknowledgements:}\\

The author thanks M. Henkel, I. Georgiev and U. T\"auber for the useful 
comments. The author thanks the access to the NIIFI Cluster-GRID, LCG-GRID 
and to the Supercomputer Center of Hungary.

\end{multicols}
\end{document}